\documentclass[12pt]{article}
\usepackage[latin1]{inputenc}
\usepackage{amsmath}
\usepackage{amssymb}
\usepackage{cite}
\makeatletter

\usepackage{a4}
\usepackage{latexsym}

\begin{document}
\title{Twist to close}
\author{D. V. Vassilevich\thanks{
e.mail:~Dmitri.Vassilevich@itp.uni-leipzig.de}\\
{\it Institut f\"{u}r Theoretische Physik, Universit\"{a}t Leipzig,}\\
{\it Postfach 100 920, D-04009 Leipzig, Germany}\\
{and}\\
{\it V.A.Fock Institute of 
Physics, St.Petersburg University, Russia}}
\maketitle

\begin{abstract}
It has been proposed that the Poincare and some other symmetries
of noncommutative field theories should be twisted. Here we extend
this idea to gauge transformations and find that twisted gauge symmetries
close for arbitrary gauge group. We also analyse twisted-invariant 
actions in noncommutative theories. 
\end{abstract}
For the development of noncommutative field theories (see reviews
\cite{NCrevs})
it was very important to recognise that the usual Poincare symmetry
must be replaced by a twisted symmetry to allow for a Lorentz-invariant
interpretation of noncommutative space-times and of the dynamics
of quantum field on them
\cite{Oeckl,Chaichian:2004za,Wess:2003da,42,Chaichian:2004yh}. Short afterwards
twisted conformal symmetries \cite{Matlock:2005zn,62}
and twisted diffeomorphism \cite{Aschieri:2005yw} were also
constructed (to mention bosonic symmetries only). Twisting the diffeomorphisms
has led to noncommutative deformations of gravity 
\cite{Aschieri:2005yw,Aschieri:2005zs,Zupnik:2005ph}.

In the standard approach to noncommutative gauge theories on cannot
close gauge groups except for $U(n)$ (or pseudo-unitary groups).
To construct gauge theory actions for other gauge groups
one has to use the Seiberg-Witten map \cite{Seiberg:1999vs}
which is a non-linear map known as an expansion in the
noncommutativity parameter. Another problem with closing the gauge
symmetries was noted recently \cite{Vassilevich:2006uv} in an
attempt to construct noncommutative counterparts to generic
2D dilaton gravities \cite{Grumiller:2002nm}.

In this paper we extend the notion of twisted symmetries to gauge
transformations of the Yang-Mills type. We show that in the twisted
case {\it any} gauge group can be closed. We also analyse invariant
noncommutative actions for a system of interacting scalars and gauge
fields. Twisted gauge invariance imposes very mild restrictions
on the form of the action (which very similar to the commutative case).

The main element in our construction, as well as that of 
\cite{Chaichian:2004za,Wess:2003da,42,Chaichian:2004yh,Matlock:2005zn,62,
Aschieri:2005yw,Aschieri:2005zs,Zupnik:2005ph},
 is the twist operator
\begin{equation}
\mathcal{F}=\exp \mathcal{P},\qquad
\mathcal{P}= \frac i2 \theta^{\mu\nu} \partial_\mu \otimes \partial_\nu \,,
\label{twist}
\end{equation}
which acts on the tensor products of functions $\phi_1 \otimes \phi_2$.
We define the multiplication map 
$\mu (\phi_1 \otimes \phi_2)=\phi_1 \cdot \phi_2$ and
use the twist operator to construct the Moyal-Weyl representation
of the star product 
\begin{equation}
\phi_1 \star \phi_2 = \mu \circ \mathcal{F} (\phi_1 \otimes \phi_2)
= \mu_\star (\phi_1 \otimes \phi_2). \label{stapro}
\end{equation} 

Consider generators $u$ of some symmetry transformations which form a Lie
algebra. If one knows the action of these transformations 
on primary fields, $\delta_u \phi = u \phi$, the action on tensor
products is defined by the coproduct $\Delta$.
In the undeformed case the coproduct is primitive,
\begin{equation}
\Delta_0 (u)=u \otimes 1 + 1 \otimes u \,,\label{Del0}
\end{equation}
and
\begin{equation}
\delta_u (\phi_1 \otimes \phi_2)=\Delta_0 (u) (\phi_1 \otimes \phi_2)
=u\phi_1 \otimes \phi_2 + \phi_1 \otimes u\phi_2 \label{duprod}
\end{equation}
satisfies the usual Leibniz rule.

In this Letter we consider a group of transformations which consists
of the global Poincare symmetry and local gauge transformations. We 
leave the action of symmetry generators on elementary fields undeformed,
but twist the coproduct (precisely as in \cite{Chaichian:2004za}
in the case of the Poincare group alone, or in \cite{Matlock:2005zn,62}
in the case of conformal symmetry):
\begin{equation}
\Delta (u) = \exp (-\mathcal{P}) \Delta_0(u) \exp ( \mathcal{P})
\label{defDel}
\end{equation}
Obviously, twisting preserves the commutation relations. Therefore,
there is no problem of closing the commutators of gauge transformations
for {\it arbitrary gauge group}. We only have to show that one can
construct enough invariants to build up field theory actions. 

Let us consider a model describing scalar fields $\varphi^a$, $a=1,\dots,p$,
which transform according to some $p$-dimensional
representation of the
gauge group with the generators $\tau_{Nb}^{a}$ (in this representations),
and gauge fields $A_\mu^N$. Let $c_{MN}^K$ be the structure
constants, $[\tau_N,\tau_M]=c_{MN}^K \tau_K$. We consider $\varphi^a$ and
$A_\mu^N$ as primary (or elementary) fields. 
The action of the gauge transformations
with local parameters $\sigma^N(x)$ (functions) is undeformed,  
\begin{equation}
\delta_\sigma \Phi = \mathcal{S} \cdot \Phi \,,\label{traPh}
\end{equation}
where
\begin{equation}
\Phi =  \left( \begin{array}{c} \varphi^a \\ A_\mu^K \\ 1 
\end{array} \right), \qquad
\mathcal{S} = \left( \begin{array}{ccc} \sigma^N \tau_{Nb}^{a} & 0 & 0 \\
0 & \sigma^M c_{MN}^K & -\partial_\mu \sigma^K \\
0 & 0 & 0 \end{array} \right).
\end{equation}
It is important that there is no star product in (\ref{traPh}).

The twisted coproduct reads in the $\theta$-expansion
\begin{eqnarray}
&&\Delta (\mathcal{S}) = \mathcal{S} \otimes 1 + 1 \otimes \mathcal{S}
-\frac i2 \theta^{\mu\nu} ([\partial_\mu ,\mathcal{S}] \otimes \partial_\nu
+ \partial_\mu \otimes [\partial_\nu ,\mathcal{S}] )\nonumber\\
&&\qquad -\frac 18 \theta^{\mu\nu}\theta^{\rho\lambda}
([\partial_\mu,[\partial_\rho,\mathcal{S}]] 
\otimes \partial_\nu \partial_\lambda + \partial_\mu \partial_\rho \otimes
[\partial_\nu,[\partial_\lambda,\mathcal{S}]]) 
+\mathcal{O}(\theta^3)\label{delS} 
\end{eqnarray}
(cf. eq.\ (3.12) of \cite{Zupnik:2005ph} for a similar expression for the
the diffeomorphism generators). Now we are ready to calculate the action of 
the gauge transformations on star-products of the elementary fields:
\begin{equation}
\delta_\sigma (\Phi^i \star \Phi^j)
=\mu_\star \circ \Delta (\mathcal{S}) (\Phi^i \otimes \Phi^j)
= S^{ik}\cdot (\Phi^k \star \Phi^j)+
 S^{jk}\cdot (\Phi^i \star \Phi^k). \label{delprod}
\end{equation}

Consider now particular invariants starting with polynomial interactions of
the scalar fields. Let
\begin{equation}
\mathcal{L}_n = d_{a_1a_2 \dots a_n} \varphi^{a_1} \star \varphi^{a_2} \star
\dots \star \varphi^{a_n} \,,\label{Ln}
\end{equation}
where $d$ is a constant tensor on the representation space of the gauge group.
\begin{equation}
\delta_\sigma  \mathcal{L}_n = \sigma^N \cdot  d_{a_1a_2 \dots a_n}
\left( \tau_{Na}^{a_1} 
\varphi^a \star \varphi^{a_2} \star \dots \varphi^{a_n} + \dots +
\tau_{Na}^{a_n} \varphi^{a_1} \star \dots \star \varphi^a \right). 
\end{equation}
The Lagrangian $\mathcal{L}_n$ is invariant under the gauge transformations
iff $ d_{a_1a_2 \dots a_n}$ satisfies the condition
\begin{equation}
 d_{aa_2 \dots a_n}\tau_{Na_1}^{a} + \dots 
+  d_{a_1a_2 \dots a}\tau_{Na_n}^{a} =0
\label{invd}
\end{equation}
for all $N$. In other words, $ d_{a_1a_2 \dots a_n}$ must be an invariant
$n$-linear form on the representation space of the gauge algebra to which 
the fields $\varphi^a$ belong. This symmetry criterion is precisely the 
same as in the commutative case (modulo the fact that $ d_{a_1a_2 \dots a_n}$
must not be symmetric under the permutations of its' indices).

Let us define the covariant derivative by the equation
\begin{equation}
\nabla_\mu \varphi = \partial_\mu \varphi + A_\mu \star \varphi \label{covder}
\end{equation}
(or, in the components, $\nabla_\mu \varphi^a = \partial_\mu \varphi^a +
A_\mu^N \tau_{Nb}^{a} \star \varphi^b$). Its' gauge transformation reads
\begin{equation}
\delta_\sigma \nabla_\mu \varphi 
= \sigma^N \tau_N \cdot (\nabla_\mu \varphi)\,. \label{dnab}
\end{equation}
The kinetic term
\begin{equation}
\mathcal{L}_{\rm kin}=
\nabla_\mu \varphi^a \star \nabla^\mu \varphi^b \eta_{ab} \label{kin}
\end{equation}
is gauge invariant iff
\begin{equation}
\tau_{Na}^{c}\eta_{cb} + \tau_{Nb}^{c} \eta_{ac}=0.\label{invkin}
\end{equation}
For example, if $\eta_{ab} = \delta_{ab}$ one gets the condition
$\tau_{Nb}^{a}=-\tau_{Na}^{b}$ meaning that the representation $\tau$
must be orthogonal. This again coincides with the commutative case.
 
The field strength is defined in the usual way as a commutator of two covariant
derivatives,
\begin{equation}
F_{\mu\nu}=\partial_\mu A_\nu - \partial_\nu A_\mu +
A_\mu \star A_\nu - A_\nu \star A_\mu \,. \label{Fmn}
\end{equation}
Note, that $F_{\mu\nu}$ does not belong to the same Lie algebra as $\tau_N$,
but, nevertheless, it is still a matrix acting in the same linear space.
The gauge transformation of $F_{\mu\nu}$ is an adjoint action,
\begin{equation}
\delta_\sigma F_{\mu\nu} =\sigma^K \cdot [\tau_K,F_{\mu\nu}], \label{dFmn}
\end{equation}
but on ${\rm Mat}_p$ rather than on the gauge algebra itself. Furthermore,
\begin{equation}
\delta_\sigma F_{\mu\nu} \star F^{\mu\nu} 
= \sigma^K \cdot [\tau_K,F_{\mu\nu}\star F^{\mu\nu} ]
\end{equation}
and
\begin{equation}
\delta_\sigma {\rm tr}( F_{\mu\nu} \star F^{\mu\nu})=0.\label{invgau}
\end{equation}

We conclude that the usual noncommutative Lagrangian
\begin{equation}
\mathcal{L}=  {\rm tr}( F_{\mu\nu} \star F^{\mu\nu})
+\mathcal{L}_{\rm kin} + \sum_n \mathcal{L}_n \label{ncL}
\end{equation}
is twisted gauge invariant
if the criteria (\ref{invd}) and (\ref{invkin}) are satisfied.

To summarise, the main idea of this approach is to apply undeformed
(``commutative'') symmetry transformations to primary (elementary) 
fields and then extend these transformations to products (tensor 
products and star-products) by using a twisted coproduct, see eq. 
(\ref{delprod}).
By the constructions, all gauge groups can be closed. One builds the
gauge invariants through star-polynomials of fields, covariant derivatives,
and field strength, like in the commutative case. The equations of motion 
are twisted gauge covariant.
 
This approach has several advantages. First of all, one does not have to invoke
nonlinear field redefinitions \cite{Seiberg:1999vs} to close the gauge 
algebra. Perhaps, even global issues of gauge theories can be handled
easier since the principal bundle remains undeformed. Physical consequences
of the twisted gauge invariance are still to be studied, as well as many 
other aspects as, e.g., extensions to non-linear realisations
(sigma-models). 

One application is more or less obvious. To couple the Munich model of
noncommutative gravity \cite{Aschieri:2005yw} to spinors, one needs local
Lorentz symmetry. Most probably, it should be a twisted gauge symmetry,
as well as the diffeomorphisms in this model.

{\bf Note added}.
After this work was completed and posted on the net I became aware that
for the first time twisting the gauge symmetries was proposed 6 years
ago by Oeckl \cite{Oeckl} without analyzing however the structure of
gauge invariants. Even earlier, it was proposed to twist physical symmetries
of field theories in \cite{KM}, though with a different twist operator.
Meanwhile, a paper \cite{Aschieri:2006ye} appeared, which develops a scheme
similar to the one presented here. The work \cite{Aschieri:2006ye}
demonstrated that the field equations derived from the first term on
the right hand side of (\ref{ncL}) are inconsistent, but this difficulty
can be resolved by considering the enveloping algebra valued gauge
fields. Additional fields decouple in the limit $\theta\to 0$, but
nevertheless their presence may have measurable consequences in cosmology.
One should note, that so far there is no complete classification 
of twisted gauge invariants (e.g., one can modify the non-linear term
in (\ref{Fmn}) without breaking the twisted gauge invariance). This
issue requires further investigations.
\subsubsection*{Acknowledgement}
This work was supported in part by the DFG project BO-1112/13-1.
I am grateful to H.~Grosse, R.~Oeckl. P.~P.~Kulish and J.~Zahn for helpful
comments on the previous version of this Letter. 

\end{document}